# Effect of substitution and heat treatment route on polycrystalline FeSe$_{0.5}$Te$_{0.5}$ Superconductors

**M. Y. Hacisalihoglu**[1,2*], **E. Yanmaz**[1]

[1]Department of Physics, Karadeniz Technical University, 61080 Trabzon, Turkey

[2]Department of Physics, Recep Tayyip Erdoğan University, 53100 Rize, Turkey

*E-mail: yasin.hacisalihoglu@rize.edu.tr

**Abstract**

The effect of atomic substitution of Te in iron based superconductors FeSe (1:1 type), which exhibits the simplest crystal structure among the iron-based superconductors, has been investigated in terms of structural, electronic transport, and magnetic properties. Polycrystalline samples with nominal Se:Te in FeSe$_{1-x}$Te$_x$ samples for x=0.5$_{350°C}$, 0.5$_{700°C}$, 0.6$_{700°C}$, and 0.75$_{700°C}$ were synthesized by the solid-state reaction method. In overall samples, it has been observed that the most superconducting properties seen in x=0.6$_{700°C}$ samples, for 0.5$_{350°C}$ and 0.5$_{700°C}$ samples latter showed better superconducting properties as dc magnetic susceptibility, ac susceptibility, and resistivity measurements and sample homogenity. $T_c$ enhancement is well correlated with the Te substitution up to 75% .

## 1. Introduction

Since the discovery of high-transition-temperature (high-$T_c$) superconductivity in layered copper oxides, extensive efforts have been devoted to the exploration of new high-$T_c$ superconducting materials. In February 2008, Hideo Hosono and coworkers reported the discovery of superconductivity in fluorine doped LaFeAsO with 26 K transition temperature has generated tremendous interest in the scientific community. [1] Shortly after in the oxypnictide system REFeAsO called the "1111" type where RE is a rare earth element, (RE= Pr, Sm, Nd, Gd, Ce, Tb, Dy), with electron- or hole-doping has produced critical temperatures
 as high as 56 K using different synthesis techniques [2-7]. In the oxygen-free compounds of AEFe$_2$As$_2$, called the "122" type where AE is an alkaline earth element (AE= Ba, Ca, Sr), with electron- or hole-doping has produced critical temperatures as high as 38 K using different synthesis techniques [8-10]. In another iron based superconductor system AFeAs, called the "111" type where A is an alkaline element (A= Li, Na ), without doping has produced critical temperatures as high as 31 K, and LiFeP showed 6 K $T_c$ without As [11-13]. Exploring for compounds that have a similar Fe-based plane like the FeAs plane resulted with a new family of iron based superconductors FeSe, called the "11" type showed $T_c$ of 8 K, which has the same iron pnictide layer structure, but without separating layers [14]. The superconducting transition temperature was enhanced to 27 K under high pressure [15]. Therefore, chemical substitutions of the ions Se-site (Te, S ) showed superconductivity up to 15 K and the Fe-site (Co, Mn, Ni, Cu, Zn) showed superconductivity up to 10 K[16, 17]. The discovery of superconductivity in $K_{1-x}Fe_{2-y}Se_2$ 30K [18] materials with the same crystal structure as $BaFe_2As_2$ has given rise to a new materials research.

According to phase diagrams of FeSe[19] and FeTe[20] parent samples superconducting tetragonal β-phase is narrow and below 450°C , and below 840°C, respectively. There are various solid state reaction heat treatment routes for Iron-chalchogenide samples varying from 300°C to 800°C [14, 21-25]. Two different heat treatment routes (one is lower and the other one is equal to 700°C) and various Se:Te substituted samples have been prepared in order to investigate the heat treatment route and substitution effect to the superconducting properties of the samples in terms of structural, electronic transport, magnetic properties, and sample homogenity.

## 2. Experimental procedure

The $FeSe_{1-x}Te_x$ sample with the nominal stoichiometric ratios of x= 0.50, 0.60, and 0.75 were prepared by the solid state reaction method. The high purity powder of Fe (99.9+%, <10 micron, Alfa Aesar), Se (99.99%, -100 mesh, Aldrich) and Te (99.997%, -30 mesh, Aldrich) were used as the precursor materials. The first step is to reduce the particle size of Te to a certain level for getting a more homogenous mixture. The granular powder of Te were ground by hand and sieved by a mesh of 63 microns. Then all three powders were mixed by hand for 30 min. in an agate mortar in the stoichiometric ratios and from each group (x= 0.50, 0.6, and 0.75), about 2.5 g powders were pressed into pellets with 650 MPa. These pellets were put into a quartz tube and sealed under a vacuum of $10^{-4}$ mbar. The calcination process was performed at 700°C for 12h. All pellets were removed from the quartz tube and ground by hand, and finally were pressed at 650 MPa, then sealed in a quartz tube and again sintered at 700°C for 22h. For comparision, we tried another heat treatment route for x=0.5 and one more sample for x=0.5 was calcined at 350° C for 12 h and sintered at 650° C for 12h, and hereafter these samples are referred to x=0.50$_{700°C}$ and x=0.5$_{350°C}$, respectively.

Structural analysis was carried out in an X-ray diffractometer (XRD), Rikagu Dmax/III. CuK$_α$ radiation was used and X-rays were generated at 40 kV and 30 mA. Measurements were performed at 2θ between 20° and 60° with steps of 0.05°. The superconducting transition temperature was determined by moment–temperature, resistivity, and ac susceptibility measurement at temperatures between 3K and 20K using a Quantum Design Physical Properties Measurement System (PPMS). The magnetic properties (M–H loops) of samples were determined using a Vibrating Sample Magnetometer (VSM) on the same system.

## 3. Results and discussion

X-ray diffraction patterns of the FeSe$_{1-x}$Te$_x$ samples for 0.5$_{350°C}$, 0.5$_{700°C}$, 0.6$_{700°C}$, and 0.75$_{700°C}$ after the sintering process are illustrated in Fig. 1. All reflections are well indexed using the tetragonal space group P4/nmm (No. 129) . In the pattern for 0.5$_{350°C}$ sample reflections from the minority impurity phase indicated by "*" refers to the hexagonal FeSe phase. For 0.5$_{700°C}$ and 0.6$_{700°C}$ sample reflections from the minority impurity phases indicated by "*", "#", "+" refers to hexagonal FeSe phase, Fe$_2$O$_3$ and Fe$_3$O$_4$, respectively. For these two samples, iron oxide phases may be related to oxygen leakage from defects occuring on the quartz tube during the heat treatment process. For the 0.75$_{700°C}$ sample, we have not observed any hexagonal impurity phases.

Observation of hexagonal impurity phases except the Te rich 0.75$_{700°C}$ sample can be explained by the phase diagrams of FeSe [19], FeTe [20] . According to these phase diagrams, aimed β-tetragonal phases are very close to the hexagonal phases and the β-FeTe tetragonal phase is more stable than the β-FeSe tetragonal phase as the sintering temperature and atomic weight percentage variation.

Figure 2 shows the (101) reflections for all samples. It has been observed that (101) reflections has shifted to the smaller angles in the $0.5_{700°C}$ sample according to $0.5_{350°C}$. Shifting can be related to the formed phases with different stoichiometric ratios after using different heat treatment routes. Reflections from $0.5_{700°C}$ and $0.6_{700°C}$ samples are very similar and according to us this is related to the formed phases with very close stoichiometric ratios after heat treatments. (101) reflections for $0.6_{700°C}$ and $0.75_{700°C}$ have been shifted to the smaller angles with an ascending Te doping content. This shifting is related to the lattice expansions due to the fact that the ionic radius of Te is larger than that of Se. This result has accordance with [16, 22].

Figure 3 presents $c$ lattice parameters were calculated with the d-spacing formula for all samples. It has been observed that $c$ lattice parameters obtained for the $0.5_{700°C}$ sample larger than $0.5_{350°C}$ sample. $c$ lattice parameters were obtained for the $0.6_{700°C}$, $0.75_{700°C}$ samples in agreement with [22] and also it has been observed that in these two samples $c$ parameters increased with increasing the Te doping content. The increasing $c$ parameters are related to the fact that the ionic radius of Te is larger than that of Se as reported in Yeh's paper [16].

Figure 4 presents the temperature dependence of the magnetic susceptibility $\chi$ ($\chi$ has been deduced from the measured magnetization and is not corrected for demagnetization effects) measured at the zero-field cooling (ZFC) mode in an 10 Oe applied field. We have determined $T_c^{mag}$ from the intersection point of two virtual lines plotted parallel to the measured magnetization line: one is before and one is after the breakpoint of the line. Using this approach we have obtained $T_c^{mag}$ to be 12.68 K, 10.88 K, 11.75 K, and 8.57 K for $0.5_{350°C}$, $0.5_{700°C}$, $0.6_{700°C}$, and $0.75_{700°C}$ samples, respectively. Below $T_c^{mag}$, it has been observed that samples have diamagnetic responses corresponding to superconductivity. The strongest superconducting signal and the best sample homogeneity has been clearly observed for x=$0.6_{700°C}$. For $0.5_{350°C}$ and $0.5_{700°C}$ samples, the low heat treated one has a higher transition temperature but poorer superconducting properties. As expected it has been observed that the x=$0.75_{700°C}$ sample has the lowest transition temperature.

The temperature dependence of a.c. magnetic susceptibility at the zero-field cooling (ZFC) mode is shown in Fig. 5. We have determined the $T_c^{ac}$ superconducting transition temperature from the intersection point of the imaginary and real part of the a.c. magnetic susceptibilities ($\chi''$, $\chi'$). Using this approach, we have obtained $T_c^{ac}$ to be 12.6, 12.5 K, 12.3 K, and 9.2 K, respectively for $0.5_{350°C}$, $0.5_{700°C}$, $0.6_{700°C}$, and $0.75_{700°C}$. The strongest superconducting signal and the most sample homogeneity is again clearly observed for the $0.6_{700°C}$ sample and the rest of the samples exhibiting a poorer superconducting transition.

By examining the peak location and the width of the peaks in the imaginary part of the a.c. susceptibility, we have been observed that the $0.6_{700°C}$ sample has the highest critical current density. According to the results the $0.5_{700°C}$ sample shown a better superconducting transition than the $0.5_{350°C}$. For all samples, the positive values of the suceptibility in the imaginary part after the transition could be due to traces of magnetic impurities like iron oxides in the sample as reported in [22].

Electrical resistance of the samples have been plotted as a function of temperature in Fig. 6. Resistance values have been normalized to its value at 25 K. $T_c^{zero}$ and $T_c^{on}$ values have been shown inset of the Fig. 6. $T_c^{on}$ values were obtained with the same criteria as in the d.c. magnetic susceptibility and the $T_c^{zero}$ values determined as the temperature that resistivity value decreased below $10^{-3}$. The $T_c^{zero}$ values was found to be 7 K, 12 K, 13.1 K and 9.5 K. As reported in [17] $T_c^{zero}$ values increased with Te substitution up to $x \leq 0.75$ for high heat treated samples. The highest $T_c^{zero}$ value has been observed as 13.1 K for $x = 0.6_{700°C}$ and the superconducting transition widths for $x = 0.6$ is sharper than the rest of the samples. For $0.5_{350°C}$ and $0.5_{700°C}$, the high heat treated sample has sharper superconducting transition width and a higher transition temperature than the low heat treated one.

The magnetic moment-magnetic field M-H curves of the samples measured below $T_c$ at 3K and above $T_c$ at 20K was shown in Fig. 7. The M-H curves revealed the existence of ferromagnetism for all samples. The opening of the M-H loops decrease with increase in magnetic field this indicates diamagnetic behavior. As reported in [26], it has been thought that diamagnetic signal has been screened by the positive moment of the magnetic impurities. The existence of ferromagnetism shows that ferromagnetism and superconductivity compete each other. According to XRD analysis the ferromagnetic background could be a result of either intrinsic trace amounts of magnetic and α-phase impurities as indicated before [27, 28]. For all samples, it has been observed that M-H loops below $T_c$ has more magnetic energy loss and less saturation magnetization than M-H loops above $T_c$. These results can be related to diamagnetic signals below $T_c$.

## 4. Conclusion

In summary, it has been studied the structural, electronic transport, and magnetic properties of the FeSe$_{1-x}$Te$_x$ samples for x = 0.5$_{350°C}$ , 0.5$_{700°C}$ , 0.6$_{700°C}$, and 0.75$_{700°C}$ superconducting materials has been studied. It has been known that atomic substitution of Te to the layered PbO-type β-FeSe modifies the superconductivity. We have observed that the x=0.6$_{700°C}$ sample showed highest $T_c^{zero}$ value with ~13.1 K and most homogenous superconducting properties. Also, $T_c$ enhancement is well correlated with the Te substitution up to 75%. We have also observed that the 0.5$_{700°C}$ sample showed higher $T_c$ values and more homogenous superconducting properties than 0.5$_{350°C}$ . According to our opinion, these results and structural similarities in 0.5$_{700°C}$ , 0.6$_{700°C}$ samples are all related to the formed phases and these phase's stoichiometric ratios after the heat treatments. Because of these reasons, precise structural analysis and phase analysis by means of EDX and SEM should be done after the heat treatment to reveal phase separation and effects of the formed phases to the superconducting properties.

To obtain more absolute results about the heat treatment routes and substitution effects on superconducting properties, it is planned to extend these measurements for various heat treatment routes and various sample preparation techniques. We also plan to investigate the formed phases various measurement techniques like EDX and SEM as a future work.

**Acknowledgements:** The authors would like to thanks to Dr. Taner Yildirim for valuable discussions.

# FIGURE CAPTIONS

Figure 1. X-ray diffraction patterns for the FeSe$_{1-x}$Te$_x$ samples after the second heat treatment.

Figure 2. The (101) reflections of the samples with various x Te doping contents.

Figure 3. Calculated *c* parameters according to the x Te doping content.

Figure 4. The temperature dependence of d.c. magnetization for with zero field cooling (ZFC) mode FeSe$_{1-x}$Te$_x$ samples.

Figure 5. The temperature dependence of a.c. magnetization for the zero field cooling (ZFC) mode FeSe$_{1-x}$Te$_x$ samples.

Figure 6. The temperature dependence of normalized resistivity for FeSe$_{1-x}$Te$_x$ samples

Figure 7. The magnetic moment vs. magnetic field measurements of FeSe$_{1-x}$Te$_x$ samples

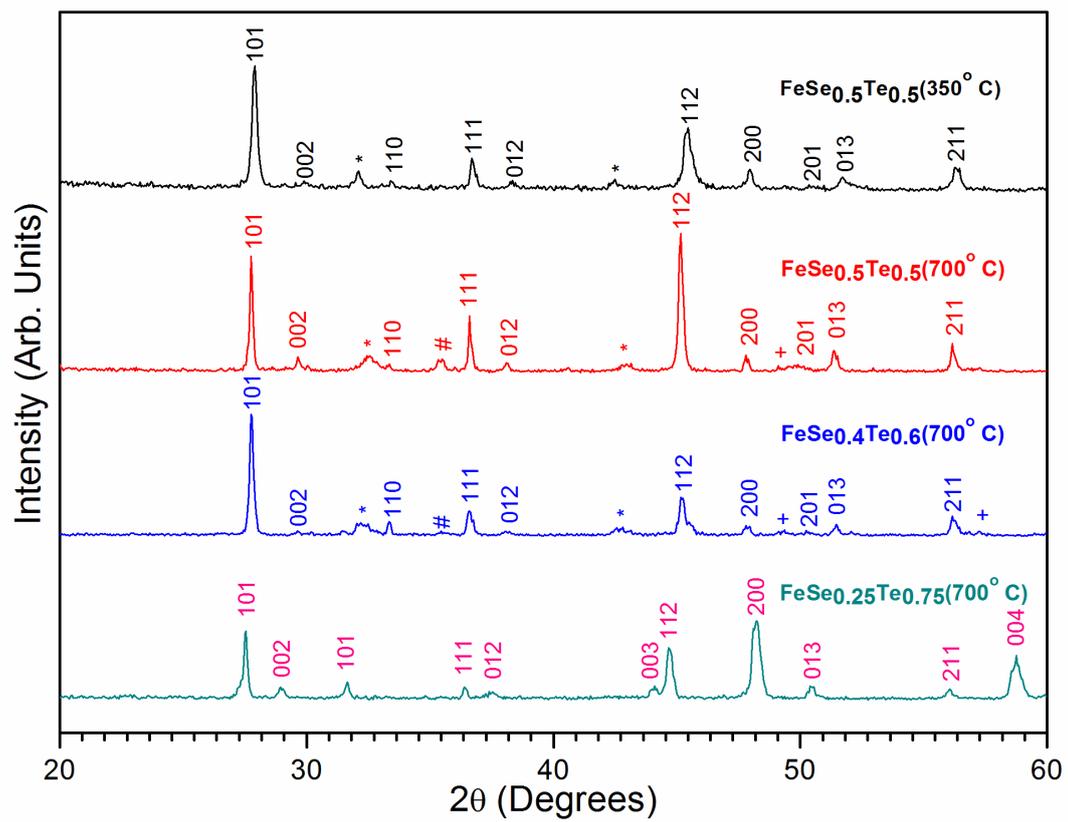

**Figure 1.**

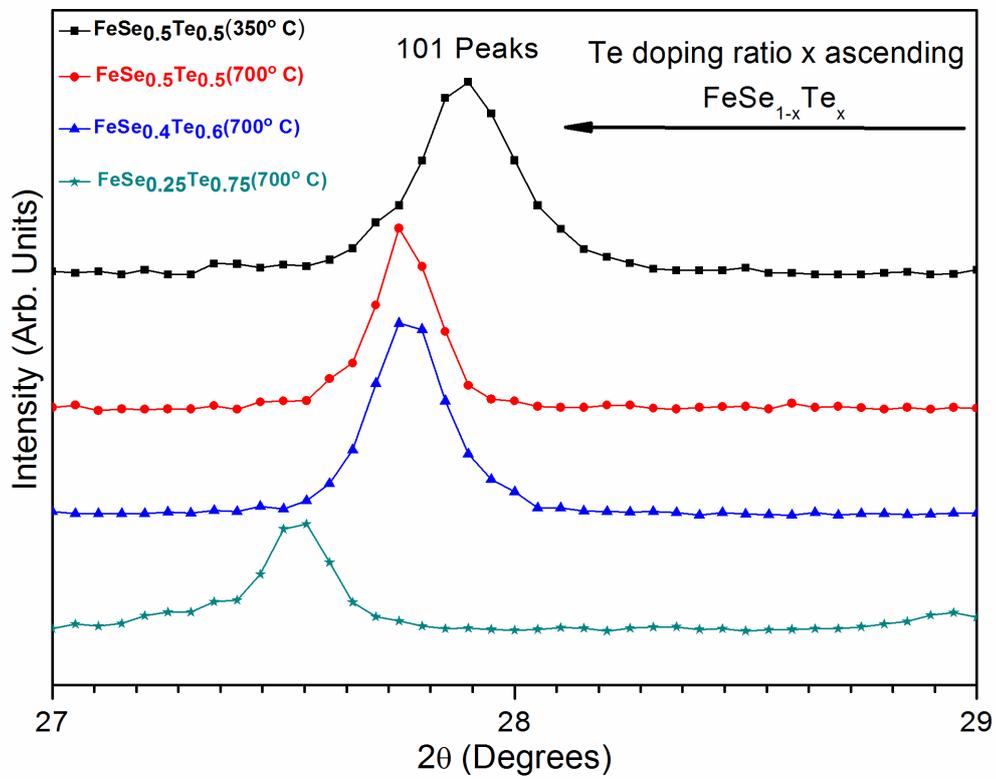

**Figure 2.**

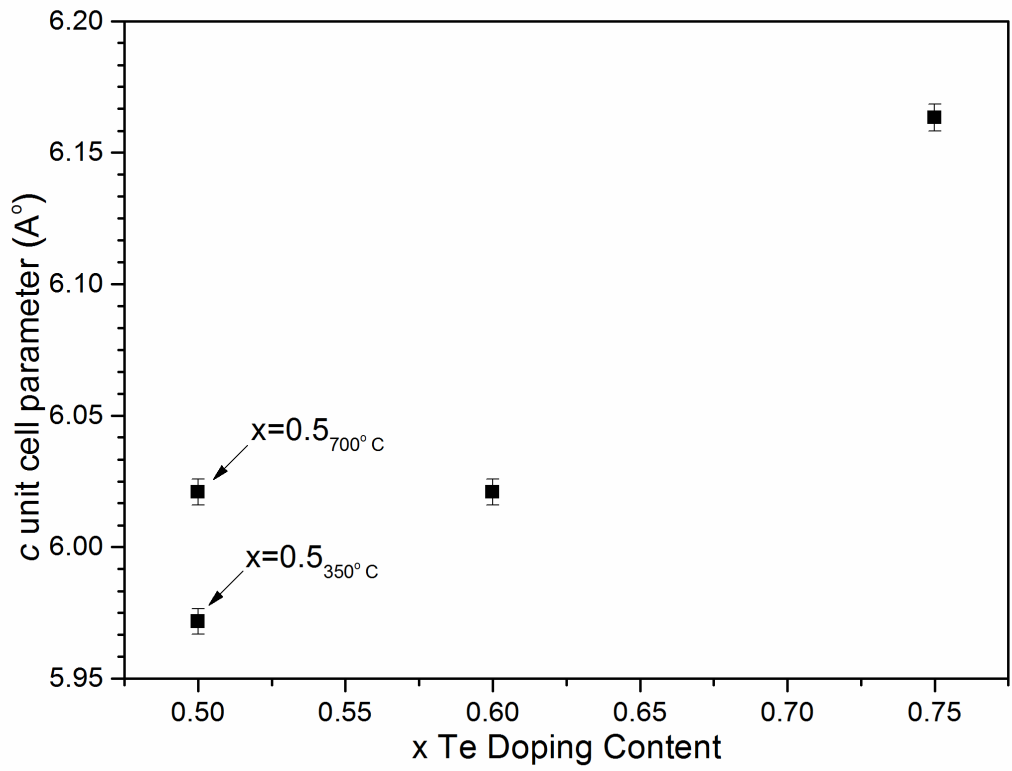

**Figure 3.**

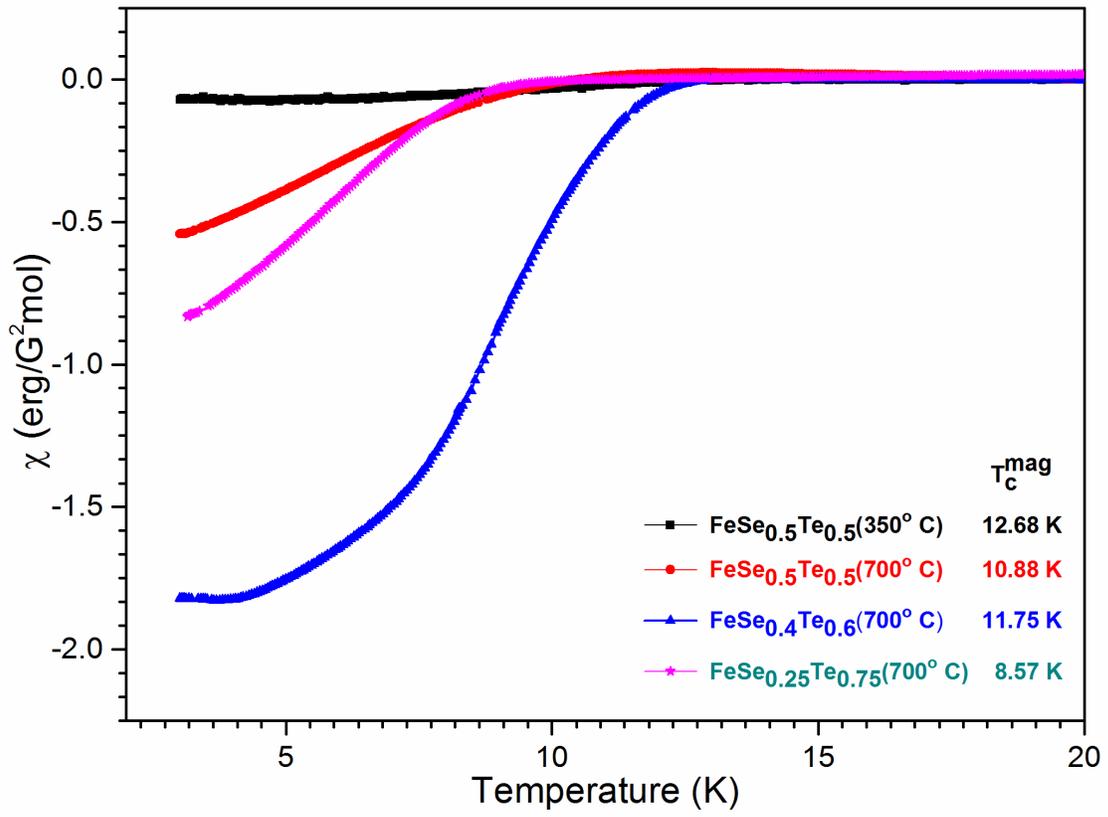

**Figure 4.**

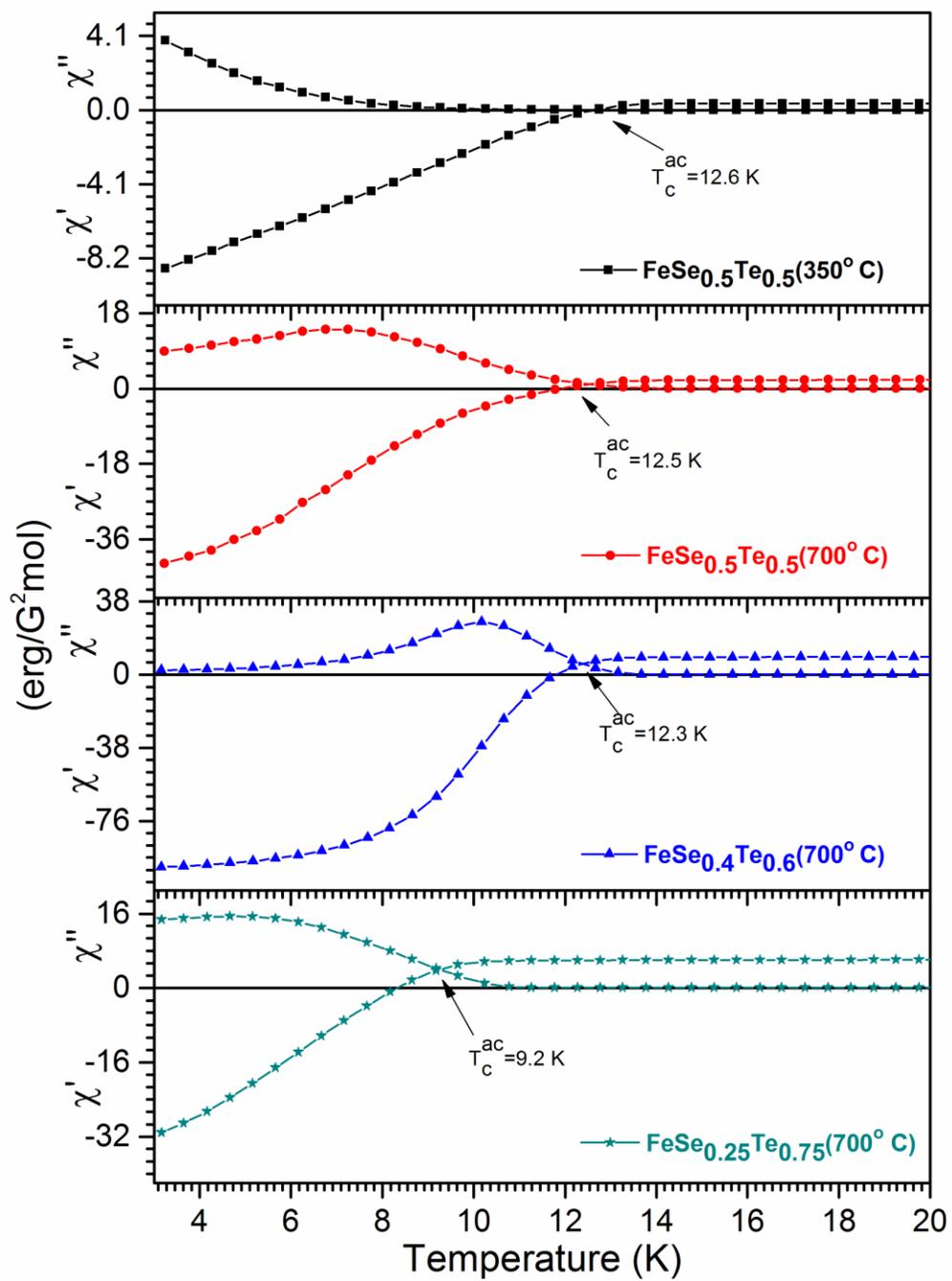

**Figure 5.**

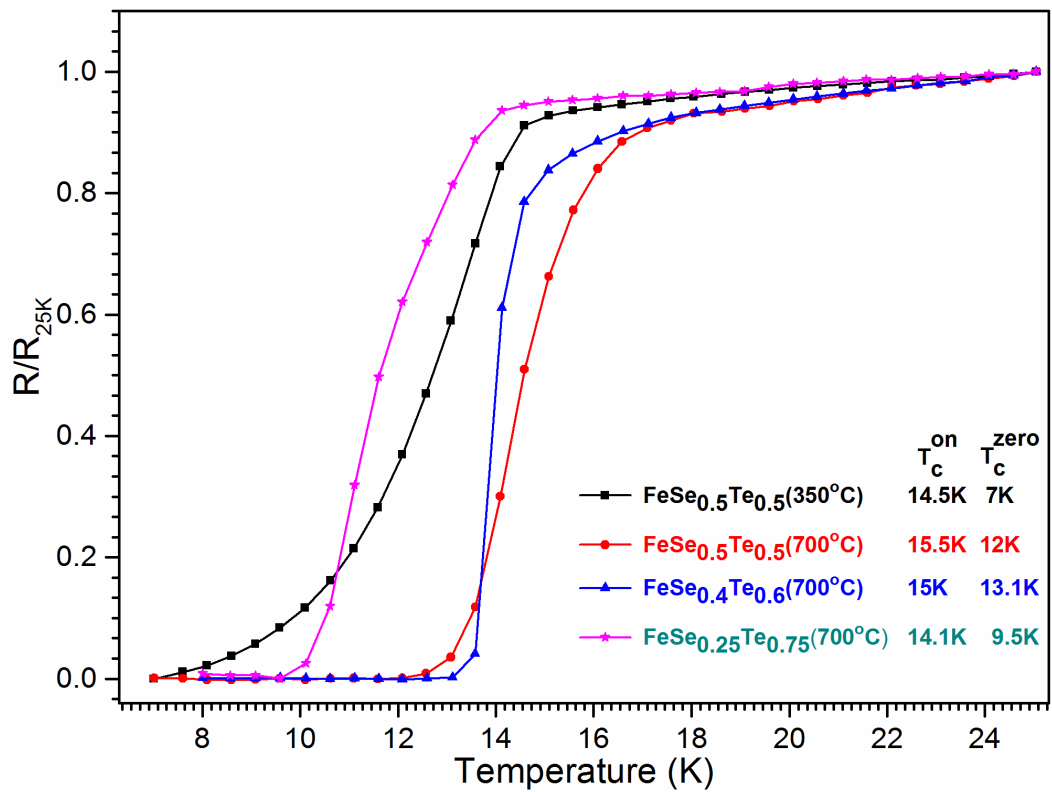

**Figure 6.**

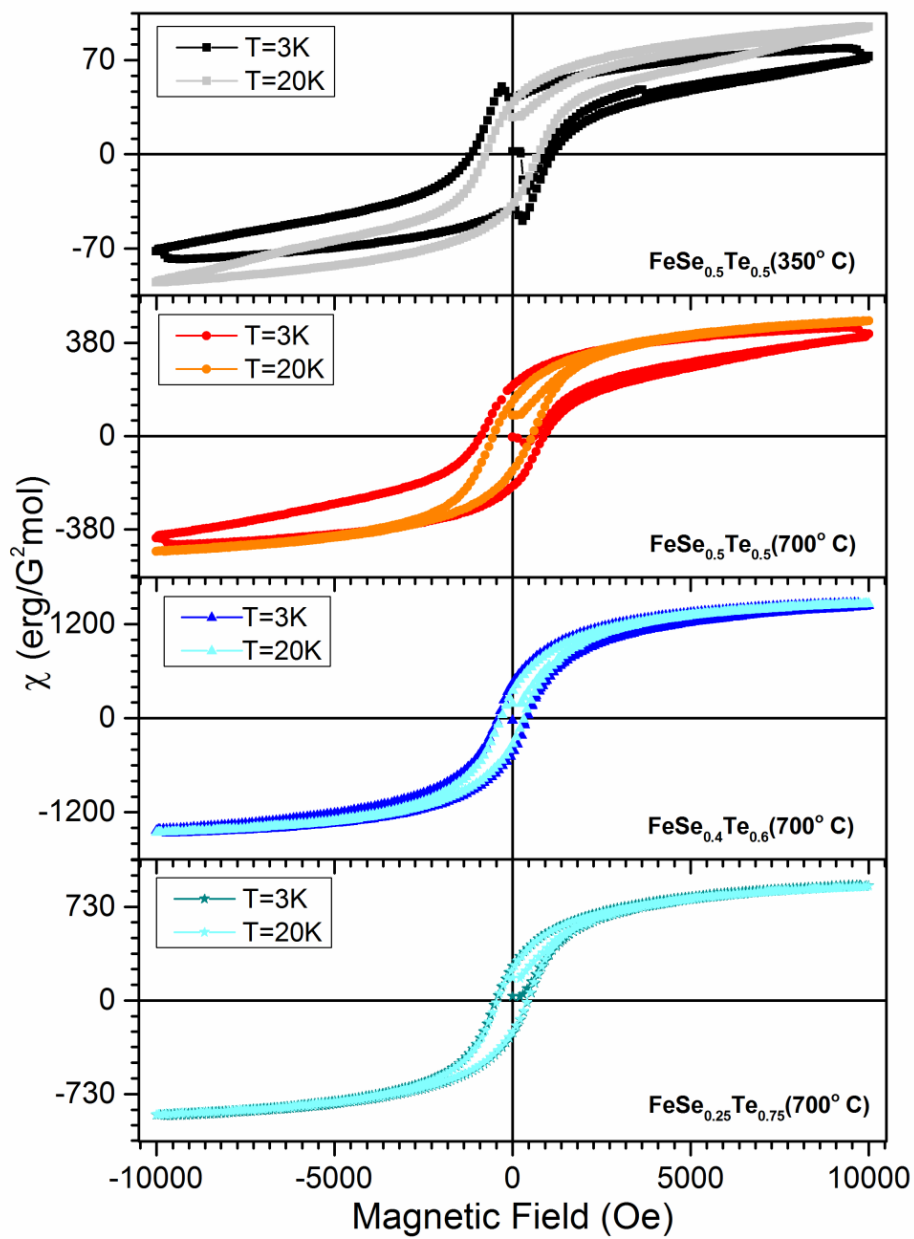

**Figure 7.**